\documentstyle[12pt]{article}
\begin{document}
\begin{center}
{\Large{\bf Statistics of Composite Fermions in Quantum Hall Effect}}
\vskip0.35cm
{\bf Keshav N. Shrivastava}
\vskip0.35cm
{\it School of Physics, University of Hyderabad,\\
Hyderabad  500 046, India}
\end{center}
\vskip0.5cm
The high Landau level filling fractions 5/2, 7/3 and 8/3 are
interpreted by using the angular momentum model. It is found that
for the odd number of flux quanta, the quasiparticles called the
``composite fermions'' are fermions but for even number-, the
quasiparticles are a mixture of bosons and fermions.Therefore,
the theory of ``composite fermions" is internally inconsistent.
\vskip5.0cm
keshav@mailaps.org   Fax. +91-40-3010145
\newpage
\baselineskip24pt
\section{Introduction}

In 1980 von Klitzing et al [1] have shown that the Hall current
is given by an integer multiple of $e^2/h$ so that an accurate
value of $e^2/h$ can be measured. Later on, it was found by Tsui
et al [2] that not only the integer but also the fractional
multiples of $e^2/h$ can be identified at higher magnetic fields
than in von Klitzing's case. The integers as well as the
fractions which multiply $e^2/h$ may be treated as spectra and
hence as quasiparticles. It was thought that these quasiparticles
obey fractional statistics [3]. The electric and the magnetic
field vectors of the Maxwell equations are decoupled so that
magnetic flux quanta are attached to the electrons which then
obey fractional statistics [4]. At this time, we found [5] the
series which gives the fractions at which quasiparticles occur.
All of the predicted fractions are the same as those observed in
various experiments. In one of the experimental measurements, it
was found that masses of two quasiparticles are equal for
different values of the fractions. Such an equality of masses can
be interpreted [6] by means of particle-hole symmetry based on
the earlier theory [5]. Several years later, flux quantization
was applied [7] to the present problem which also, gave
fractions. One of the series of fractions obtained from this work
is the same as in our paper [5] and there is one more series which is slightly
shifed from our's. Laughlin [8] wrote the wave function for the
quasiparticle of a fractional charge. In this theory, the vacuum state is
correctly defined from which the quasiparticles of fractional
charge are created and when sufficient number of particles are created
they may be annihilated subject to the condition that the flux area
is held rigid so that the quasiparticles are incompressible and hence
the hardest objects in the world which GaAs is not. Our work predicts
 the masses in terms of
particle-hole symmetry by using the Kramers theorem [9,10].
\vskip0.15cm
\begin{center}
{\bf Table 1: } Interpretation of 5/2, 7/3 and 8/3 in terms of
theory of ref. 5.
\vskip0.25cm
\begin{tabular}{ccccc}
\hline
$l$ & $l/(2l+1)$ & $(l+1)/(2l+1)$ & $(nl/(2l+1)$ & $n(l+1)/(2l+1)$\\
\hline
$\infty$ & 1/2  & 1/2  & 5/2 & 5/2\\
        7 & 7/15 & 8/15 & 7/3 & 8/3\\
\hline
\end{tabular}
\end{center}
\vskip0.5cm
In this communication, we interpret the fractions 7/3, 5/2 and
8/3 seen in the measurements. We
report that 7/3 and 8/3 are particle-hole conjugates and 5/2 is
the $n=5$ state of the level at 1/2. Since we used angular
momentum theory to intepret the quantum Hall effect we are able
to attach spin to the effective fractional charges found in the
data which leads to identification of statistics. The odd fluxes
are always fermions whereas even number of fluxes are mixtures of
fermions and bosons. Therefore the ``composite fermion" theory of
Jain is incorrect. 

\section{Fractions 7/3, 5/2 and 8/3}

Our model [5] has several features which are in agreement with
experimental measurements. The fractions predicted by us are the
same as those experimentally found by Eisenstein and St\"ormer
[11], Willett et al [12] Du et al [13] and others. The fractions
occur in two groups. One group of fractions lies on high field
side of 1/2 and the other on the low field side of 1/2. The value
of 1/2 occurs in both the series and hence represents a fluid of
two components. The grouping of fractions in our theory is the
same as experimentally found [11]. Our model also gives $\nu=1/2$
for very large values of $l$ and hence there is a limiting value
$n/2$ where $n$ is the Landau level quantum number. In our model,
one of the series is,
\begin{equation}
\nu = {l+(1/2)-s\over 2l+1} = {l\over2l+1}
\end{equation}
which predicts one group of fractions 0, 1/3, 2/5, 4/9, 5/11,
etc. which are also observed by Willett et al [12]. Another group
of fractions is predicted [5] by the expression,
\begin{equation}
\nu = {l+({1\over2})+s\over2l+1} = {l+1\over2l+1}
\end{equation}
which are 1, 2/3, 3/5, 4/7, etc in complete agreement with the
experimental data [11-13].  When $l=\infty$ both the above series
approach 1/2 except that one series approaches from the right
hand side and the other from the left hand side exactly as
observed [11].  The left and the right side approaches arise from
the Kramers conjugate states and the predicted approach is
exactly as observed. Due to the limit, there is a Fermi surface
at 1/2. However, we can shift from the Fermi surface to higher
values when higher Landau levels are occupied. The fraction 1/2
becomes $n/2$ with $n$ as the Landau level quantum number. Thus
1/2, 3/2, 5/2, 7/2, 9/2, etc., become allowed. This predicted
feature with an odd numerator with 2 in the denominator is also
exactly as observed. Thus for $n=1$, we have two series, one
merging from left while the other merging from right at 1/2 and
the same picture is repeated for different values of $n$. The
entire pattern of pairwise series is observed exactly as
predicted. Many fractions with 2 in the denominator have been
observed by Lilly et al [14] and by Yeh et al [15]. The series
(1) and (2) above can be used to explain the higher Landau levels
easily. Eisenstein et al [16] have found that at higher values of
the Landau level quantum number, $n$, the number of fractions
observed is much less than at the lowest Landau level. At the
magnetic field of 4 to 5 Tesla only a small number of fractions
are observed, the strongest ones are at 8/3, 5/2 and 7/3. Since
there is a charge versus Landau level quantum number product, it
is not possible to distinguish large charge from a large Landau
level quantum number. In the angular momentum series, $l/(2l+1)$ 
is the particle-hole conjugate of $(l+1)/(2l+1)$ by virtue of
Kramers theorem. For $l=7$ two values, 7/15 and 8/15 are
predicted and $l=\infty$ values are $\pm1/2$. When the same particle
occurs in different levels, its charge remains unchanged. We can
multiply the values by $n=5$ so that the predicted values of 1/2,
7/15 and 8/15 become 5/2, 7/3 and 8/3. These predicted values are
exactly the same as those observed experimentally by Eisenstein
et al[16]. Thus 7/3 is the particle-hole conjugate of 8/3 as seen in
Table 1 for $n=5$. Here the particle-hole symmetry is obtained by
reversing the spin, -1/2 for one and +1/2 for the other as in
Kramers conjugate pairs due to strong spin-orbit interaction
whereas in ordinary semiconductors the particle has the same spin
as the hole and they are separated in energy by a gap in the
range of optical frequencies. Thus the present problem is
different from the usual optical absorption in semiconductors.
The present spin-orbit interaction has $1/c$ in the coupling constant
and both spin orientations are permited to determine the gyromagnetic
ratios whereas the ordinary spin-orbit interaction has $1/c^2$ and
the gyromagnetic ratio is determined by the Lande's formula.
Thus the angular momenta series (1) and (2) given by ref. 5
explain the quantum Hall effect correctly.

\section{Composite Fermions}

We introduce the flux quantization such that the field inside the
material is smaller than outside. The reduction in the field is
linearly proportional to the electron density, the number of
electrons per unit area, $\rho$. We use the even number of
fluxons, $2\phi_o$ or $2p\phi_o$ where $p$ is an integer so that
the effective field becomes,
\begin{equation}
B^* = B - 2p\rho\phi_o\,\,\,.
\end{equation}
The quasiparticles which experience the field $B^*$ are called
composite fermions, CFs, and the field $B$ is experinced by the
usual electrons. As far as the flux quantization is concerned the
fractions are completely symmetric, 
\begin{equation}
\nu = {\rho\phi_o\over B}
\end{equation}
for electrons and
\begin{equation}
\nu^* = {\rho\phi_o\over B^*}
\end{equation}
for the CFs. We substitute (3) in (5) and then (4) in the
resulting equation so that
\begin{equation}
\nu_+ = {\nu^*\over 2p\nu^*+1}.
\end{equation}
For $\nu^*=0,1,2,3,\cdots$ etc. we can get $\nu=0, 1/3, 2/5, 3/7$, 
etc. When we reverse the sign of the field shift by reversing the
sign of $2p\rho\phi_o$, the series generated is,
\begin{equation}
\nu_- = {-\nu^*\over 2p\nu^*-1}
\end{equation}
which gives, $\nu_-=0, 1, 2/3, 3/5, 4/7, \cdots$, etc. Both of
these series are the same as those of Shrivastava [5] except that
the $\nu_-$ series is shifted in the value of $\nu^*$. Because of
this shift, there is a serious loss because (1) and (2) addup to
one but (6) and (7) do not. If both series in (6) and (7) are
correct, when we put $p=1$, the effective field of (3) becomes
$B^*=B-2\rho\phi_o$ so that two (even number) of fluxes, $2\phi_o$,
is multiplied to the areal density of electrons [7]. For $p=2,
4\phi_o$ and for $p=3, 6\phi_o$ are needed. Here $\nu=1/3$ means
that effective charge is $(1/3)e$, etc. However, the quantity
which enters the definition of $\nu$ is the area multiplied by
charge. Hence we can not know whether the area or the charge has
to become 1/3. The fractions $\nu$ for electrons and $\nu^*$ for
CFs are interchangable in (6) and (7), 
\begin{equation}
\nu^* = {\nu\over 2p\nu+1}
\end{equation}
and
\begin{equation}
\nu^* = {-\nu\over 2p\nu-1}.
\end{equation}
By comparing (6) with (8) and (7) with (9), we find that $\nu$
and $\nu^*$ are completely interchangable. One of the CF series
(6) is identical to (1) and the other series (7) can be made
equal to (2) if
\begin{equation}
{l+1\over2l+1} =  {\nu^*\over 2p\nu^*-1}
\end{equation}
The solution of which is
\begin{equation}
\nu^* = l+1\,\,.
\end{equation}
This is an important result because it brings CF series in
agreement with those obtained by use of the angular momentum [5].
The series in (1) is,
\begin{equation}
{l+{1\over2}-s\over2l+1} = {l\over2l+1}
\end{equation}
for $s={1\over2}$ and that in (2) is $[l+{1\over2}+s]/(2l+1)$ which for
$s={1\over2}$ becomes $(l+1)/(2l+1)$. Therefore (11) is
equivalent to
\begin{equation}
l+{1\over2} + s = \nu^*\,\,\,.
\end{equation}
Substituting this relation in (5) and writing $\rho=n_o/A$, the
number of electrons $n_o$ in the area $A$, we obtain,
\begin{equation}
l +{1\over2} + s = {n_o\phi_o\over AB^*}\,\,.
\end{equation}
For $l=0, s={1\over2}$, the above gives 
\begin{equation}
AB^* = n_o\phi_o
\end{equation}
where $n_o$ can be even or odd. Thus both the even as well as old
values of $n_o$ correspond to fermion statistics due to spin,
$s={1\over2}$. When $l=0, s=0,$ the result (14) gives
\begin{equation}
AB^* = 2n_o\phi_o
\end{equation}
which means that bosons $(s=0)$ are having even number of flux
quanta. Thus CFs are bosons for even number of $\phi_o$ and
fermions for both even and odd number of flux quanta. Therefore,
for odd number of flux quanta, the quasi particles are fermions
and for even number, they are mixtures of bosons and fermions.
The field $B^*$ is zero at,
\begin{equation}
B^*=B-2p\rho\phi_o = {\rho\phi_o\over\nu} - 2p\rho\phi_o=0
\end{equation}
so that
\begin{equation}
{1\over\nu} = 2p
\end{equation}
which for $p=1$, gives $\nu={1\over2}$. Therefore, for half
filled Landau level there is a possiblity of zero resistivity but
the experimental zeros of $\rho_{xx}$ are not located at
$\nu=1/2$. If the flux is reversed, $\nu=-1/2$ occurs which does
not have a physical interpretation in terms of band filling.
Therefore, Jain's formula of $B^*$ given by (3) is not in
agreement with the experimental data of $\rho_{xx}$. Therefore,
attachment of $2\phi_o$, even number of fluxes is not consistent
with the data. We require that $\rho_{xx}$ should touch zero at
$\nu=1/3$ but that is not born out from the value of $B^*$ which
attaches even number of flux quanta, when $\nu=1/3, p=3/2$ but
$p$ is required to be an integer. Therefore, the electric field
(resistivity) is not understood if even fluxes are attached to
electrons, when $\nu=1/3, p=3/2$, the effective field becomes,
\begin{equation}
B^*=B-3\rho\phi_o
\end{equation}
so that odd number of flux quanta, $3\phi_o$ are attached. This
creates the problem of statistics. If we put $\nu^*=1/3$ in (13),
quasiparticles of different statistics must be mixed.\\
     It is true that Laughlin's theory [8] of fractional charge
has one-body interaction summed over all the particles from $1$ to $N$
and a factor which has two-body interaction and the product
is made from $1$ to $N$. This forms the many body wave 
function using which the matrix elements of the hamiltonian containing
electron-electron repulsive interaction, electron-nuclear attraction
and respective kinetic energies, are computed to find the energy.
The charge of $1/3$ is assumed while writing the two-body correlation.
This value of the charge is taken from the experimental measurements
of the plateau in the Hall resistivity. On the other hand, our theory 
of ref.[5] is surely a single-particle theory and gives the correct
charge. At $l\to\infty$, we get $\nu=1/2$ for both (1) and (2). At this
point our energy is $-(1/2)l$ for $j=l+1/2$ and $+(1/2)(l+1)$
for $j=l-1/2$ so that the energy diverges at $l\to \infty$, $\nu=1/2$.
The interaction of the form $\hat{n}\times\vec{v}$.$\vec{s}$ where
 $\hat{n}$
is a unit vector in the direction of $\vec{r}$, $\vec{v}$ the velocity and
 $\vec{s}$
the spin, has a distance dependent coupling constant, $f(|\vec{r}|)$ so that
$-f(|\vec{r}|)\vec{l}.\vec{s}$ depends on the distance. We can introduce the summation
over $N$ particles to make it into  many-body hamiltonian but it is not required
to get the fractional charge. Similarly, for one electron per unit area, the flux is
quantized without the need for many-body interactions.

{\it Inconsistencies}. It is agreed that even number of flux quanta when 
attached
 to an
electron give ``composite fermions" which must be {\it fermions}. Similarly, 
odd number of flux quanta attached to the electron are called ``composite
bosons" and they are {\it bosons}. In eq.(15) integer number of flux quanta
 give fermions and in eq.(16) even number of flux quanta give bosons.
Therefore, there are inconsistencies in the Jain's theory of composite
 fermions. In the composite fermions spin is completely arbitrary. Therefore,
 at $\nu=1/2$ four states, one singlet and three components of the triplet
should occur. On the other hand $\nu=1/3$ is found to be polarized
and $\nu=1/2$ gives only two states. This experimental observation
 is consistent with the theory of Shrivastava [5] and not with
 that of the composite fermion theory of Jain [7]. The claim
of Melinte et al [17] that the composite fermion theory agrees
 with the data of polarization of electrons in the NMR experiment is
therefore incorrect.

It was found[10] that a lot of data on the quantum Hall effect is in fact
in accord with the theory of Shrivastava [5]  and the theoretical Table 1 
of Shrivastava [5] is the same as the experimental Fig.18 of St\"ormer's
Nobel lecture[18]. The experimentally found high Landau levels are in
 agreement with our theory [19]. We find that there is a change in the
 magnetic moment of the electron [20] due to the new way of looking at the
 spin-orbit interaction. There is an important  effect of sweep rate on
the quantum Hall effect resistivity minima[21]. The fact that our
 predicted results are in complete agreement with the data is
clearly brought out in a recent article [22].

In conclusion, we are able to understand the high Landau levels
by using the theory of angular momentum given in ref. 5. We find
that the composite fermions are not able to explain the zeroes in
the transverse resistivity. By means of flux quantization, the
odd number of fluxes give Fermi statistics but even number of
fluxes are obtained for both the Fermi as well as Bose
statistics. Therefore, the CF theory of Jain [7]  is internally inconsistent. 
\newpage
\noindent{\bf References}
\begin{enumerate}
\item K. von Klitzing, G. Dorda and M. Pepper, Phys. Rev. Lett.
	{\bf45}, 494 (1980).
\item D.C. Tsui, H.L. St\"ormer and A.C. Gossard, Phys. Rev. Lett.
	{\bf48}, 1559 (1982).
\item F. Wilczek, Phys. Rev. Lett. {\bf48}, 957 (1982).
\item D. Arovas, F. Wilczek and J.R. Schrieffer, Phys. Rev. Lett.
{\bf53}, 722 (1984).
\item K.N. Shrivastava, Phys. Lett. A{\bf113}, 435 (1986);
{\bf115}, 459(E) (1986).
\item K.N. Shrivastava, Mod. Phys. Lett. {\bf13}, 1087 (1999).
\item J.K. Jain, Phys. Rev. Lett. {\bf63}, 199 (1989). This paper
has the same series as in ref. 5.
\item R.B. Laughlin, Phys. Rev. Lett. {\bf50}, 1395 (1983).
\item K.N. Shrivastava, Superconductivity: Elementary Topics,
World Scientific, Singapore, 2000.
\item K.N. Shrivastava, CERN  SCAN-0103007.
\item J.P. Eisenstein and H.L. St\"ormer, Science {\bf248}, 1510 (1990).
\item R. Willett, J.P. Eisenstein, H.L. Stormer, D.C. Tsui, A.C.
Gossard and J.H. English, Phys. Rev. Lett. {\bf59}, 1776 (1987).
\item R.R. Du, H.L. St\"omer, D.C. Tsui, L.N. Pfeiffer and K.W.
West, Phys. Rev. Lett. {\bf70}, 2944 (1993).
\item M.P. Lilly, K.B. Cooper, J.P. Eisenstein, L.N. Pfeiffer and
K.W. West, Phys. Rev. Lett. {\bf83}, 824 (1999).
\item A.S. Yeh, H.L. St\"ormer, D.C. Tsui, L.N. Pfeiffer, K.W.
Baldwin and K.W. West, Phys. Rev. Lett. {\bf62}, 592 (1999).
\item J.P. Eisenstein, M.P. Lilly, K.B. Cooper, L.N. Pfeiffer and
K.W. West,  Physica E{\bf 6}, 29 (2000).
\item S. Melinte, N. Freytag, M. Horvatic, C. Berthier, L. P. Levy,
V. Bayot and M. Shayegan, Phys. Rev. Lett.{\bf84}, 354 (2000).
\item H. L. St\"ormer, Rev. Mod. Phys. {\bf71},875(1999), Nobel lecture.
\item K. N. Shrivastava, Mod. Phys. Lett. B {\bf14},1009 (2000)[cond-mat/0103604].
\item K. N. Shrivastava, cond-mat/0104004.
\item K. N. Shrivastava, cond-mat/0104577.
\item K. N. Shrivastava, in {\it Frontiers of Physics}, edited by
B. G. Sidharth, Kluwer Press, New York  2001.
\end{enumerate}

\end{document}